\documentclass[11pt]{article}
\usepackage{moriond,epsfig}

\def\Journal#1#2#3#4{{#1} {\bf #2}, #3 (#4)}

\def\PLB{{\em Phys. Lett.} B}
\def\PRL{\em Phys. Rev. Lett.}
\def\PRD{{\em Phys. Rev.} D}

\def\be{\begin{equation}}
\def\ee{\end{equation}}
\def\bea{\begin{eqnarray}}
\def\eea{\end{eqnarray}}

\def\lsim{\mathrel{\mathop
  {\hbox{\lower0.5ex\hbox{$\sim$}\kern-0.8em\lower-0.7ex\hbox{$<$}}}}}
\def\gsim{\mathrel{\mathop
  {\hbox{\lower0.5ex\hbox{$\sim$}\kern-0.8em\lower-0.7ex\hbox{$>$}}}}}
\def\ii{\'{\i}}

\begin{document}
\vspace*{4cm}
\title{TACHYONIC PREHEATING AND SPONTANEOUS SYMMETRY BREAKING}

\author{ J. GARC\'IA-BELLIDO }

\address{Theory Division, C.E.R.N., CH-1211 Gen\`eve 23, Switzerland}

\maketitle 

\abstracts{We discuss the recent scenario of tachyonic preheating at
  the end of inflation as a consequence of a tachyonic mass term in
  the scalar field responsible for spontaneous symmetry breaking. We
  use 3D lattice simulations to expore this very non-perturbative and
  non-linear phenomenon, which occurs due to the spinodal instability
  of the scalar field. Tachyonic preheating is so efficient that
  symmetry breaking typically completes within a single oscillation of
  the field distribution as it rolls towards the minimum of its
  effective potential.}

During the last few years we have learned that the coherent
oscillations of a scalar field may induce explosive particle
production within a dozen oscillations, due to a nonperturbative
process called {\em preheating}~\cite{KLS}. Usually preheating is
associated with broad parametric resonance in the presence of a
coherently oscillating inflaton field~\cite{KLS}, but other mechanisms
are also possible. In a recent letter~\cite{FGBKLT} we studied what we
called {\it tachyonic preheating}, which occurs due to the spinodal
instabilities in the scalar field responsible for symmetry breaking.
Spontaneous symmetry breaking is one of the fundamental ingredients of
modern theories of elementary particle physics. In the context of the
evolution of the universe it has often been considered as associated
with first or second order {\em thermal} phase transitions. We
explored a new scenario~\cite{FGBKLT} in which symmetry breaking
occurs at zero temperature, at the end of a period of inflation, when
the tachyonic mass term $-m^2\phi^2/2$ appears suddenly, i.e. on a
time scale that is much shorter than the time required for symmetry
breaking to occur, and which induces the spinodal growth of quantum
fluctuations. Spontaneous symmetry breaking is a strongly nonlinear
and nonperturbative effect. It usually leads to the production of
particles with large occupation numbers inversely proportional to the
coupling constants~\cite{FGBKLT}. As a result, the perturbative
description, including the Hartree and $1/N$ approximations, has
limited applicability. For instance, it cannot describe adecuately the
rescattering of the created particles and other important features
such as production of topological defects. Thus, for further
theoretical understanding of these issues one should go beyond
perturbation theory. This is the reason why we used the new methods
of lattice simulations, based on the observation that quantum states
of bose fields with large occupation numbers can be interpreted as
classical waves and their dynamics can be fully analyzed by solving
relativistic wave equations on a lattice~\cite{lattice,FT}. A
significant advantage of these methods as compared to other lattice
simulations of quantum processes is that the semi-classical nature of
the effects under investigation allows us to have a clear visual
picture of all the processes involved.

We will show that tachyonic preheating can be extremely efficient,
both in the usual symmetry breaking model and in hybrid
models~\cite{Hybrid}.  In most cases it leads to the transfer of the
initial potential energy density $V(0)$ into the gradient or kinetic
energy of scalar particles within a single oscillation. For instance,
contrary to standard expectations, the first stage of preheating in
hybrid inflation~\cite{GBL} is typically tachyonic, which means
that the stage of oscillations of a homogeneous component of the
scalar fields driving inflation either does not exist at all or ends
after a single oscillation. A detailed description of our results will
be given in a coming publication~\cite{FGBKLT2}.

Symmetry breaking occurs due to tachyonic instability and may be
accompanied by the formation of topological defects. Here we will
consider two toy models that are prototypes for many interesting
applications, including symmetry breaking in hybrid inflation.
The simplest model of spontaneous symmetry breaking is based on the
theory with effective potential
\begin{equation}\label{aB1}
V (\phi) = {\lambda\over 4}(\phi^2-{  v}^2)^2 \equiv  V(0) - {m^2\over
2}\phi^2 + {\lambda\over 4} \phi^4 \ ,
\end{equation}
where $\lambda \ll 1$. $V(\phi)$ has a minimum at $\phi = \pm v$
(or $|\phi|=v$ in the case of a complex symmetry breaking field),
and a maximum at $\phi = 0$ with negative curvature $V'' = -m^2$. 

The development of tachyonic instability in this model depends on
the initial conditions. We will assume that initially the symmetry
is completely restored so that the field $\phi$ does not have any
homogeneous component, i.e. $\langle \phi \rangle= 0$. But then,
because of the symmetry of the potential, 
$\langle \phi \rangle$ remains zero at all later stages, and for
the investigation of spontaneous symmetry breaking one needs to
find the spatial distribution of the field $\phi(x,t)$.  To
avoid this complication, many authors assume that there is a small
but finite initial homogeneous background field $\phi(t)$, and
even smaller quantum fluctuations $\delta\phi(x,t)$ that grow on
top of it. This approximation may provide some interesting
information, but quite often it is inadequate. In particular, it
does not describe the creation of topological defects, which, as
we will see, is not a small nonperturbative correction but an
important part of the problem.

For definiteness, we suppose that in the symmetric phase, $\phi=0$,
there are the usual quantum fluctuations of a massive field with mode
functions ${1 \over \sqrt{2 k_0}}e^{-ik_0t +i{\vec k \vec x}}$, where
$k_0^2=k^2+V''$, and then at $t = 0$ we `switch on' the tachyonic term
$-m^2\phi^2/2$. The modes with $k = |{\vec k}| < m$ will grow
exponentially, $\phi_k \sim \exp (t\sqrt{m^2- k^2})$, so the
dispersion of these fluctuations can be estimated as
\begin{equation}
\langle \phi^2 \rangle
 = {1 \over 4\pi^2 } \int_0^m  dk\, k \, e^{2t\sqrt{m^2- k^2}} =
{e^{2mt}(2mt-1)+1 \over 16\pi^2t^2}  \ .
\label{aBB}
\end{equation}

To get a qualitative understanding of the process of spontaneous
symmetry breaking, instead of many growing waves with momenta $k < m$
let us consider a single sinusoidal wave $\delta\phi = A(t) \cos kx$
with $k \sim m$ and with initial amplitude $A(0) \sim {m\over 2\pi}$
in one-dimensional space. This amplitude grows exponentially until its
becomes $A(t_*) \sim v = m/\sqrt \lambda$, which leads to the
splitting of the universe into domains of size ${\cal O}(m^{-1})$, in
which the scalar field changes from ${\cal O}(v)$ to ${\cal O}(-v)$.
The gradient energy density of domain walls separating regions with
positive and negative $\phi$ will be $\sim k^2\delta\phi^2= {\cal
  O}(m^4/\lambda)$. This energy is of the same order as the total
initial potential energy of the field $V(0) = m^4/4\lambda$.  This is
one of the reasons why any approximation based on perturbation theory
and ignoring topological defect production cannot give a correct
description of the process of spontaneous symmetry breaking.  Thus a
substantial part of the false vacuum energy $V(0)$ is transferred to
the gradient energy of the field $\phi$ when it rolls down to the
minimum of $V(\phi)$. Because the initial state contains many quantum
fluctuations with different phases growing at a different rate, the
resulting field distribution is very complicated, so it cannot give
all of its gradient energy back and return to its initial state $\phi
= 0$.  This is could be the main reason why spontaneous symmetry
breaking and the initial stage of preheating in this model may occur
within a single oscillation of the field~$\phi$.

Consider the tachyonic growth of all fluctuations with $k < m$, i.e.
those that contribute most to $\langle \phi^2 \rangle$ in Eq.~(\ref{aBB}).
This growth continues until $\langle \phi^2 \rangle^{1/2} \sim v/2$,
since at $\phi = v/\sqrt 3$ the curvature of the effective potential
vanishes and instead of tachyonic growth one has the usual
oscillations of all the modes. This happens within the time $t_* \sim
{1\over 2 m} \ln{\pi^2\over \lambda}$. The exponential growth of
fluctuations up to that moment can be interpreted as the growth of the
occupation number of particles with $k \ll m$. These occupation
numbers at the time $t_*$ grow up to $n_k \sim \exp(2mt_*) \sim
\exp(\ln\pi^2/\lambda) = {\pi^2\over \lambda} \gg 1$. One can easily
verify that $t_*$ depends only logarithmically on the choice of the
initial distribution of quantum fluctuations. For small $\lambda$ the
fluctuations with $k \ll m$ have very large occupation numbers, and
therefore they can be interpreted as classical waves of the field
$\phi$. The dominant contribution to $\langle \phi^2 \rangle$ in Eq.
(\ref{aBB}) at the moment $t_*$ is given by the modes with wavelength
$l_* \sim 2\pi k_*^{-1} \sim \sqrt 2 \pi m^{-1} \ln^{1/2}
{(2\pi^2/\lambda)} > m^{-1}$. As a result, at the moment when the
fluctuations of the field $\phi$ reach the minimum of the effective
potential, $\langle\phi^2\rangle^{1/2} \sim v$, the field
distribution looks rather homogeneous on a scale $l \lsim l_*$. On
average, one still has $\langle \phi\rangle = 0$. This implies that
the universe becomes divided into domains with two different types of
spontaneous symmetry breaking, $\phi \sim \pm v$. The typical size of
each domain is $l_*/2 \sim {\pi\over \sqrt 2 } \ m^{-1}
\ln^{1/2}{2\pi^2\over \lambda}$, which differs only logarithmically
from our estimate $m^{-1}$. At later stages the domains grow
in size and percolate (eat each other up), and spontaneous symmetry
breaking becomes established on a macroscopic scale.

\begin{figure}[ht]
\centering 
\psfig{figure=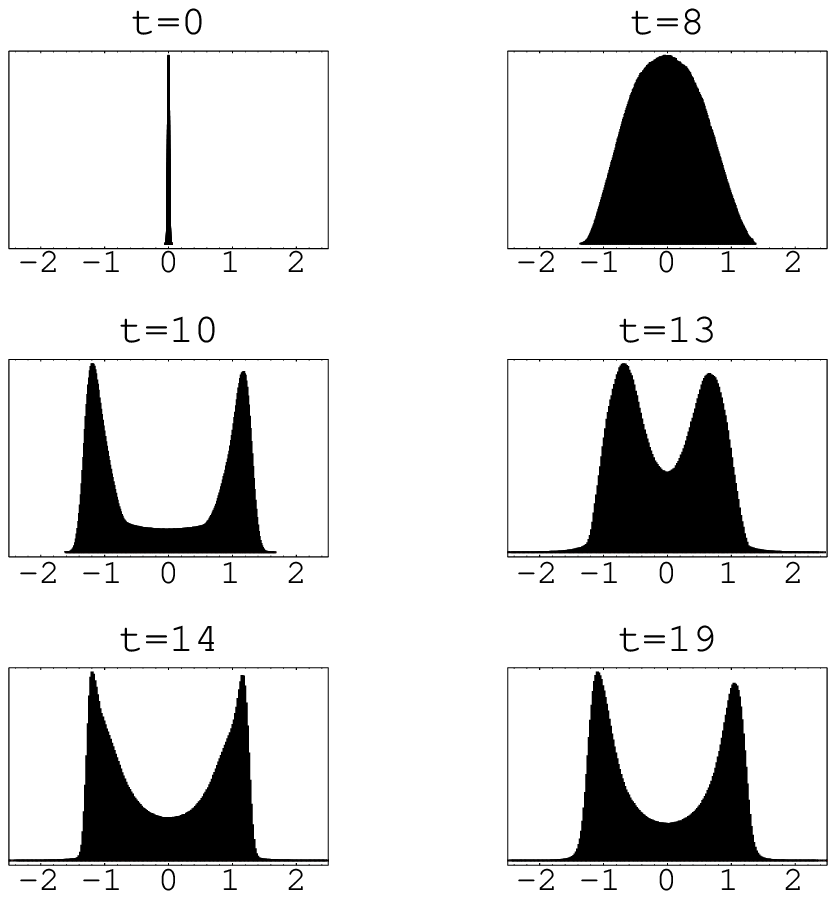,height=5.5cm}
\psfig{figure=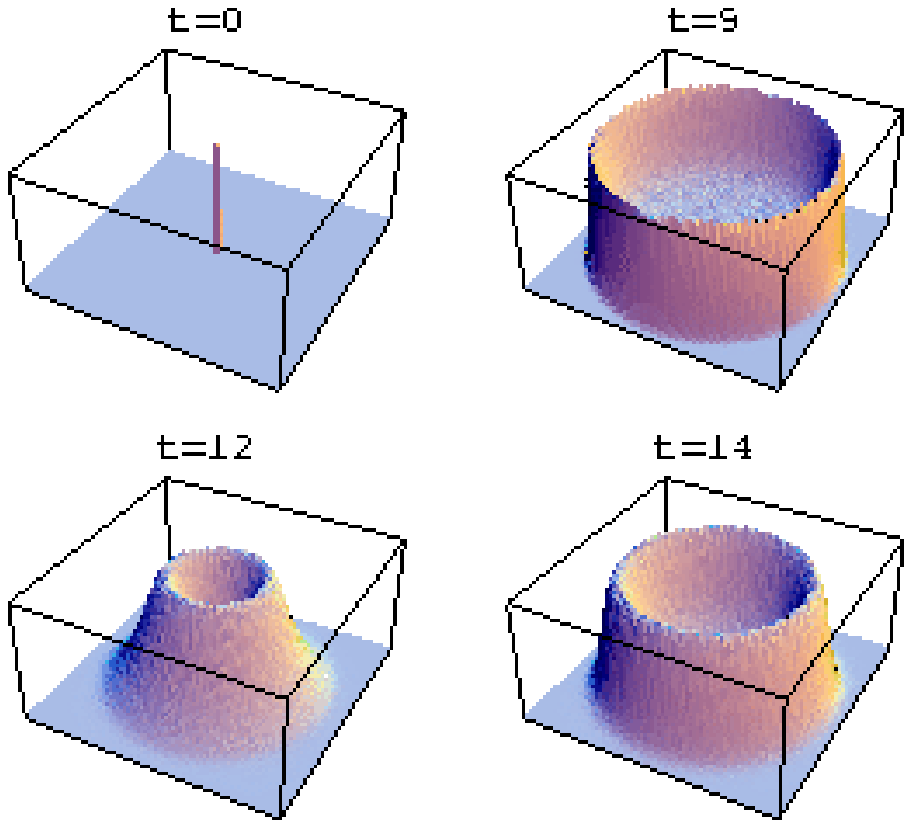,height=5.5cm}\\
\caption{\label{onefielddistrib} Left panel: The process of symmetry
  breaking in the model (\ref{aB1}) for $\lambda = 10^{-4}$. In the
  beginning the distribution is very narrow. Then it spreads out and
  shows two maxima which oscillate about $\phi = \pm v$ with an
  amplitude much smaller than $v$. These maxima never come close to
  the initial point $\phi = 0$. The values of the field are shown in
  units of $v$. Right panel: The process of symmetry breaking in the model
  (\ref{aB1}) for a complex field $\phi$. The field distribution falls
  down to the minimum of the effective potential at $|\phi| = v$ and
  experiences only small oscillations with rapidly decreasing
  amplitude $|\Delta\phi| \ll v$.}
\end{figure}

When the field rolls down to the minimum of its effective potential,
its fluctuations scatter off each other as classical waves. It is
difficult to study this process analytically, but fortunately one can
do it numerically using the method of lattice simulations developed in
Refs.~[3,4]. We performed our simulations on lattices with either
$128^3$ and $256^3$ gridpoints. Figure~\ref{onefielddistrib}
illustrates the dynamics of symmetry breaking in the model
(\ref{aB1}). It shows the probability distribution $P(\phi,t)$, that
is the fraction of the volume containing the field $\phi$ at a time
$t$ if at $t = 0$ one begins with the probability distribution
concentrated near $\phi = 0$, with the quantum mechanical dispersion
$\langle\phi^2\rangle = m^2/4\pi^2$ as in (\ref{aBB}).

As we see from this figure, after the first oscillation the
probability distribution $P(\phi,t)$ becomes narrowly concentrated
near the two minima of the effective potential corresponding to
$\phi = \pm v$. In this sense one can say that symmetry breaking
completes within one oscillation. To demonstrate that this is not
a strong coupling effect, we show the results for the model
(\ref{aB1}) with $\lambda = 10^{-4}$. Note that only when the
distribution stabilizes and the domains become large can one use
the standard language of perturbation theory describing scalar
particles as excitations on a (locally) homogeneous background.
That is why the use of the nonperturbative approach based on
lattice simulations was so important for our investigation.

The dynamics of spontaneous symmetry breaking in this model is better
illustrated by a computer generated movie that can be found at
http://physics.stanford.edu/gfelder/hybrid/1.gif. It consists of an
animated sequence of images similar to the one shown in Fig.
\ref{onefielddistrib}. These images show the whole process of
spontaneous symmetry breaking from the growth of small Gaussian
fluctuations of the field $\phi$ to the creation of domains with $\phi
= \pm v$. Similar results can be obtained for the theory of a complex
scalar field $\phi$ with the potential (\ref{aB1}).  For example, the
behavior of the probability distribution $P(\phi_1,\phi_2,t)$ in the
theory of a complex scalar field $\phi = (\phi_1 + i\phi_2)/\sqrt 2$
is also shown in Fig.~\ref{onefielddistrib}. As we can see, after a
single oscillation this probability distribution has stabilized at
$|\phi| \sim v$. A computer generated movie illustrating this process
can also be found at http://physics.stanford.edu/gfelder/hybrid/2.gif.
We also performed 3D lattice simulations of symmetry breaking in
hybrid models of inflation and found that, contrary to original
expectations, symmetry breaking also occurs within a single
oscillation, thus making tachyonic preheating a generic feature of
potentials with a negative curvature direction in the 
potential~\cite{FGBKLT}.

In summary, the new 3D lattice methods developed during the
last few years in application to the theory of reheating after
inflation have been applied to the theory of spontaneous symmetry
breaking. These methods have for the first time allowed us not only
to calculate correlation functions and spectra of produced particles,
but to actually {\it see}\, the process of spontaneous symmetry
breaking and to reveal some of its rather unexpected features, like
production of topological defects, percolation of domains, and 
thermalization.

\section*{Acknowledgments}
It is a pleasure to thank my collaborators, Gary Felder, Patrick
Greene, Lev Kofman, Andrei Linde and Igor Tkachev for many insights
into the problem of symmetry breaking. I also aknowledge support by a
NATO Linkage Grant 975389 and a Spanish CICYT grant FPA2000-0980. 
The author is on leave from the Universidad Aut\'onoma de Madrid.

\end{document}